\documentclass[12pt]{article}
\usepackage{amsmath}
\usepackage{caption}
\usepackage{hyperref}
\usepackage{authblk}
\usepackage{epstopdf}
\usepackage{cite}
\usepackage{float}
\usepackage{mathrsfs}
\usepackage{subcaption}
\usepackage{amssymb}
\usepackage{mathtools}
\usepackage{graphicx}
\DeclareMathOperator{\sech}{sech}
\usepackage{xcolor}
\usepackage{filecontents}
\usepackage[T1]{fontenc}

\definecolor{tab:red}{RGB}{200,0,0}
\definecolor{tab:blue}{RGB}{10,0,200}
\definecolor{tab:green}{RGB}{10,200,10}
\definecolor{tab:black}{cmyk}{0,0,0,1}
\definecolor{pink}{rgb}{1.0,0.75,0.8}
\definecolor{purple(x11)}{rgb}{0.41,0.21,0.61}
\definecolor{orange(webcolor)}{rgb}{0.98,0.6,0.01}
\definecolor{gray}{rgb}{0.5,0.5,0.5}
\definecolor{brown(traditional)}{rgb}{0.59,0.29,0.0}

\begin{document}
\nocite{*}
\baselineskip=18 pt
\title{Study of exponential wormhole metric in $f(R)$ gravity}
\author[1,a]{\small  Partha Pratim Nath}
\author[2,a]{\small  Debojit Sarma}

\affil[a]{\footnotesize Department of physics, Cotton University}
\affil[1]{\textit{phy1891005\_partha@cottonuniversity.ac.in}}
\affil[2]{\textit{sarma.debojit@gmail.com}}
\maketitle

\begin{abstract}
  In this work, we have studied an "exponential form" of spacetime metric: 
\begin{equation*}
ds^2 = -e^{-\frac{2m}{r}}dt^2 +e^{\frac{2m}{r}}dr^2 + e^{\frac{2m}{r}}[r^2 d\theta^2 + r^2 \sin^2\theta d\phi^2]
\end{equation*}
in some of the viable $f(R)$ gravity models, viz. exponential gravity model, Starobinsky gravity model, Tsujikawa model and Gogoi-Goswami f(R) gravity model. Here we have calculated the parameters including energy density, tangential and radial pressure for these corresponding models of $f(R)$ gravity. Subsequently we have investigated the energy conditions viz. null energy condition(NEC), weak energy condition(WEC) and strong energy condition(SEC) for the considered models. We have also explained the suitable conditions of energy for these models by related plots.  
\end{abstract}

\smallskip
\noindent \textbf{Keywords.} Wormhole Geometry, Modified gravity, Energy Conditions.

\section{Introduction}
\par Wormholes in General Relativity belong to a special class of solutions to the Einstein's Field Equations which act as tube-like or bridge-like that connects two distinct points of the same space-time of two different universes. The tubular structure is considered to be asymptotically flat on both sides. One of the main features of wormhole is the wormhole throat, which can be defined as a two dimensional hypersurface of minimal area or the point where the radius is minimum \cite{visser1997geometric, hochberg1997geometric}. The concept of this bridge-like structure was first constructed by Einstein and Rosen which is known as Einstein-Rosen bridge \cite{einstein1935particle}. They inspected the exact solution that describe the geometry of the bridge. Their solution is linked with the work of Ludwig Flamm \cite{flamm1916beitrage, flamm2015republication}, who for the first time constructed the isometric embedding of Schwarzschild solution but his solutions sustained some stability problems. Hermann Weyl in 1928 \cite{weyl1929gravitation, weyl1928zusammenhang} proposed a wormhole hypothesis of matter in connection with mass analysis of electromagnetic field theory. However he used the term "one-dimensional tube" instead of the term "wormhole". Later Ellis \cite{ellis1973ether} gave another term for wormhole known as "Drainhole". Then Wheeler \cite{wheeler1955geons} named them as "Geons" and predicted the shape of the wormhole which offers a twofold space. Wheeler and Misner \cite{misner1957classical} coined the term "wormhole" and later his solutions were transformed into Euclidean wormholes by Hawking \cite{hawking1988wormholes} and others. These theoretical objects lead to various static and non-static in proportion to the fixed or variable radius of the wormhole throat. Shortly afterward, Kar \cite{kar1994evolving} discussed the static wormhole and inquired into their properties with examples. Kar and Sahdev \cite{kar1996evolving} explored evolving Lorentzian wormholes. Kar wormholes have a quantum structure and connect different points of the space on the Planck Scale. All these wormholes were not stable and traversable.
\par Traversability is also an important feature of wormhole. If anything that enters through one side of the wormhole can exit through the other, the wormhole is traversable. In order to become traversable, the wormhole should not contain a horizon, because the presence of the horizon would prevent the two-way travel through the wormhole. Morris-Thorne \cite{morris1988wormholes} gave the idea of traversable wormhole with some new concepts such as throat. They examined the static spherically symmetric wormholes by using the principles of General Relativity and introduced the fundamental theory for traversable wormhole. The energy-momentum tensor of the matter supporting such geometries, the wormhole throat necessitates the introduction of exotic matter \cite{morris19881wormholes, visser1995lorentzian}, which leads to the violation of the null energy conditions(NEC) and the averaged null energy conditions(ANEC) \cite{visser1997geometric, morris1988wormholes, morris19881wormholes, hochberg1997self, friedman1993topological, hochberg1998null, hochberg1998dynamic} near the throat region. Exotic matter is a form of dark energy(having an EoS with $\omega<-\frac{1}{3}$),produces a repulsion. Recent observations have shown that the dark energy is solely responsible for the accelerated expansion of the universe. After that wormholes have been studied from various aspects and conditions \cite{hawking1992chronology, frolov1993wormhole, lemos2008plane, rahaman2006thin, rahaman2009thin, kim2001exact, lobo2007exotic, roman1993inflating, andino2020wormholes, sorokhaibam2020traversable, kuhfittig2020traversable, santos2021casimir}. Since the exotic matter is a troublesome issue and thus many justifications have been presented in favor of the violation of the energy conditions such as invoke quantum fields in curved spacetime, scalar-tensor theories \cite{bronnikov2004conformal, bronnikov2018scalar, shaikh2016wormholes} and so on. Many efforts have been made to reduce the use of exotic matter. "Volume integral quantifier" is one of the most famous proposition which quantifies the total amount of energy condition violating matter \cite{visser2003traversable, kar2004quantifying}. Further Nandi  and others \cite{nandi2004volume} improved this formulation to know the exact quantity of the exotic matter present in the given spacetime. Additionally, there have been proposals regarding confinement of exotic matter at the throat of the wormhole, viz. cut and paste method \cite{visser1989traversable, visser19891traversable, lobo2003linearized}. To avoid the energy violations, thin-shell wormholes \cite{poisson1995thin} were studied, where ordinary matter is concentrated on the throat of the wormhole. In Recent years, wormhole solutions are developed using the background modified gravity theories such as Kaluza-Klein gravity \cite{dzhunushaliev1999wormholes, de2009static},Born-Infeld theory \cite{richarte2009wormholes}, Brans-Dicke theory \cite{agnese1995wormholes, lobo2010general, nandi1997brans, sushkov2011composite, eiroa2008thin, papantonopoulos2020wormhole}, mimetic theories \cite{myrzakulov2016static}, $f(R)$ gravity \cite{eiroa2016thin, garcia2010wormhole, garcia2011nonminimal}, Einstein-Gauss-Bonnet theory \cite{richarte2007thin, kanti2011wormholes, mehdizadeh2015einstein, antoniou2020novel, maeda2008static}, Einstein-Cartan theory \cite{bronnikov2015wormholes, bronnikov2016wormholes, mehdizadeh2017einstein, mehdizadeh2017dynamic, di2017spin}. It has been shown in modified theories of gravity that the matter inside the wormhole may satisfy the necessary energy conditions but the effective stress-energy tensor \cite{kocuper2017stress} containing higher order derivatives is responsible for the violation of the Null Enenrgy Condition(NEC).
\par The modified or extended theories of gravity have proposed some logical explanation to some observational phenomena that can be hardly explained through the General Theory of Relativity. For example, dark energy \cite{riess2004type, peebles2003rev}, dark matter \cite{aprile2012dark, akerib2017results}, massive pulsars \cite{demorest2010two, antoniadis2013massive}, super-Chandrasekhar white dwarfs \cite{nugent2011supernova, das2013new} etc can be explained with the help gravity theories such as $f(R)$ \cite{de2010f}, $f(\tau)$ \cite{yang2011conformal, cai2016f}. Here $R$ and $\tau$ being, respectively, the Ricci and torsion scalars. One of the easiest modification of the Einstein Hilbert action is the $f(R)$ theory of gravity, in which the curvature scalar or Ricci scalar $R$ in gravitational action is replaced by $f(R)$, which is an arbitrary function of the Ricci scalar $R$ \cite{nojiri2011unified}. Buchdahl \cite{buchdahl1970non} in 1970, first proposed the $f(R)$ gravity model.
\par The field equations obtained in $f(R)$ theory are very complicated and have a larger set of solutions than that of General Theory of Relativity. Bertolami \cite{bertolami2007extra} and others simplified this theory which provides a coupling between matter and the function $f(R)$ that leads towards an extra force which may explain the acceleration of the universe \cite{bertolami2010accelerated, nojiri2007introduction}. By using $f(R)=R+\alpha R^2$, Starobinsky \cite{starobinsky1980new} first derived early inflamatory universe solution long before the effectuality of the inflamaton was known. The late time cosmic acceleration has been explained by Carroll et al \cite{carroll2004cosmic, nojiri2017modified}. in the contex of $f(R)$ gravity. Many researchers have studied numerous viable cosmological models in $f(R)$ gravity \cite{cognola2005one, bergliaffa2006constraining, capozziello2006cosmological, amarzguioui2006cosmological, santos2007energy, amendola2007f, ananda2008evolution, carloni2008evolution}. Bertplami, Sotiriou, Harko and others explored the coupling of an arbitrary function of $R$ with the matter Lagrangian density. Limiting from strong lensing, $f(R)$ gravity was studied in Patalini formalism by Yang and Chen \cite{yang2009f}. Capozziello and Laurentis \cite{capozziello2012dark} had given a different approach to dark matter problem in the context of $f(R)$ gravity. Bronnikov and Starobinsky \cite{bronnikov2007no} demonstrated that wormholes can not be formed in dark matter models governed by scalar tensor theory even in the presence of electric and magnetic field. Bronnikov et al. \cite{bronnikov2010notes} has showed that for $\frac{df}{dR}=F(R)<0$, the non-existence of wormhole could be violated in $f(R)$ theory of gravity. Both Brans-Dicke theory and $f(R)$ gravity were considered to obtain wormhole. It is shown \cite{bronnikov201016} that no vacuum wormmhole exists in Brans-Dicke theory but exists in $f(R)$ gravity if it satisfies an extremum where the effective gravitational constant changes its sign. Bronnikov et al. \cite{bronnikov2017wormholes} have shown a no-go theorem in General Relativity for obtaining wormhole solutions, according to which it eliminates the existene of wormholes with flat or AdS asymptotic regions on both sides of the throat where the source matter is isotropic.
\par In $f(R)$ gravity, Lobo and Oliveira \cite{lobo2009wormhole} wormhole solutions where the matter threading the wormhole is a fluid which satisfies the energy conditions, the higher order terms of $f(R)$ theory gives rise the energy violations. Beato et al. \cite{ayon2016analytic, canfora2017topologically} have shown that exotic matter is not mandatory for constructing traversable wormhole. Similarly Harko et al. \cite{harko2013modified}, Pavlovic and Sossich \cite{pavlovic2015wormholes} also shown that the $f(R)$ theory can describe the wormhole geometry without any kind of exotic matter. Mazharimousavi and Halilsoy \cite{mazharimousavi2016wormhole} also constructed traversable wormhole model that satisfies energy conditions. The exact solutions of traversable wormholes with non-constant Ricci scalar have been obtained by Golchin and Mehdizadeh \cite{golchin2019quasi}. On the other hand, Restuccia and Tello-Ortiz \cite{restuccia2020new} have given new class of $f(R)$ gravity model and studied cosmological parameters. Spherically symmetric Lorentzian wormholes \cite{barros2018wormhole} also have been investigated with a constant scalar curvature in quardatic $f(R)$ gravity. De Benedictis and Horvat \cite{debenedictis2012wormhole} showed the existence of wormhole throat in $f(R)$ gravity and also studied their porperties. On the other hand Sharif and Zahra \cite{sharif2013static} investigated the wormhole solutions for isotropic, anisotropic fluids and barotropic equation of state with the radial pressure. In numerous $f(R)$ models, wormholes have been studied using Karmarkar condition \cite{kuhfittig2019spherically, shamir2020traversable, fayyaz2020morris}. Many physicist studied wormholes in various $f(R)$ gravity models using different redshift and shape functions \cite{godani2019traversable, kuhfittig2018wormholes, jahromi2018static, cataldo2016static, bahamonde2016cosmological, rahaman2016finslerian}. Vittorio et.al. developed some astrophysical techniques to detect wormholes and in the same time to reconstruct the solution once they have been observed \cite{de2020general, de2021testing, de2021reconstructing, de2021epicyclic}.
\par Here in this work, we investigate the so called exponential wormhole metric in $f(R)$ gravity. For more than 60 years, this metric has been investigated by many researchers. This metric has some charming properties that it passes almost all of the standard lowest order weak field test of General Relativity, but strong field actions and medium field actions are very different. The paper is assembled as follows. In Sec.II, we have studied the exponential wormhole metric in General Relativity. Where we have studied various properties such as throat radius, Karmarkar condition, field equations, flare-out condition, Ricci convergence condition etc. In Sec.III, we have studied the exponential wormhole metric in modified $f(R)$ gravity model. First we have constructed the field equations and with the help of these we studied the exponential wormhole metric in four viable $f(R)$ gravity model. Finally we present results and discussion in the Sec.IV. 
\section{Exponential wormhole metric in General Relativity}
The exponential wormhole metric \cite{yilmaz1958new, yilmaz1973new, misner1999yilmaz, robertson1999x, ben2007relativistic, ben2011some, boonserm2018exponential},
\begin{equation}\label{(expo)}
ds^2 = -e^{-\frac{2m}{r}}dt^2 +e^{\frac{2m}{r}}dr^2 + e^{\frac{2m}{r}}[r^2 d\theta^2 + r^2 \sin^2\theta d\phi^2],
\end{equation}
has an attractive feature in weak fields \cite{boonserm2018exponential}, that is when $\frac{2m}{r}<<1$, we have
\begin{equation}
ds^2=[-dt^2+dr^2+r^2(d\theta^2 + \sin^2\theta d\phi^2)]+\frac{2m}{r}[dt^2+dr^2+r^2(d\theta^2 + \sin^2\theta d\phi^2)].
\end{equation}
i.e.,
\begin{equation}
g_{ab}=\eta_{ab} +\frac{2m}{r} \delta_{ab}.
\end{equation}
As this matches with the lowest order field expansion, so the exponential metric will pass all the lowest order weak field test of General Relativity. But strong field and medium field behaviour are very different \cite{boonserm2018exponential}.
\subsection{Wormhole throat}
In order to find the radius of the throat of the wormhole, let us consider the area of the spherical surface,
\begin{equation}
S(r)= 4\pi r^2 e^{\frac{2m}{r}}.
\end{equation} 
For the extremum value of $S(r)$,
\begin{equation}
\frac{d S(r)}{dr}=8\pi (r-m)e^{\frac{2m}{r}}=0,
\end{equation}
which gives $r=m$, which is the radius of the throat. Because at $r=m$, we find that
\begin{equation}
\frac{d^2 S(r)}{dr^2}= 8\pi e^2>0,
\end{equation}
i.e., the area has a minimum at $r=m$. Again, all the metric components are finite and the diagonal components are non-zero at the throat (i.e. at $r=m$).
\subsection{Karmarkar Condition}
For a static and spherically symmetric line element to be class one, Karmarkar \cite{karmarkar1948gravitational} developed a mandatory condition. For the exponential metric, some of Riemann curvature tensors are,
\begin{align}
\begin{split}
{}& R_{1414}= \frac{2m(m-r)e^{-\frac{2m}{r}}}{r^4} ; R_{1212}=-\frac{me^{\frac{2m}{r}}}{r}; R_{1224}=R_{1334}=0;\\
& R_{3434}=-\frac{m(m-r)\sin^2 \theta e^{-\frac{2m}{r}}}{r^2}; R_{2323}=-m(m-2r)\sin^2 \theta e^{-\frac{2m}{r}},
\end{split}
\end{align}
These Riemann components fulfilling the well known Karmarkar relation,
\begin{equation*}
R_{1414}=\frac{R_{1212}R_{3434}+R_{1224}R_{1334}}{R_{2323}}
\end{equation*}
with $R_{2323}\neq 0$. The spacetime, that satisfies the Karmarkar condition is known as embedding class one. Now, by substituting the non-zero Riemann components in the above relation, we get
\begin{equation}
\frac{m^2(2m-3r)(m-r)\sin^2\theta}{r^4}=0.
\end{equation}
Solving the above relation, we will get $m=0$ or $m=r$ or $m=\frac{3}{2}r$. We have found that the throat radius is $r=m$. That means the exponential wormhole metric fulfills the requirement of class one at or near the throat. But $m=0$ is forbidden due to the flare out condition.
\subsection{Field equations}
\begin{figure}
\centering
\begin{subfigure}[b]{0.4\textwidth}
\includegraphics[width=\textwidth]{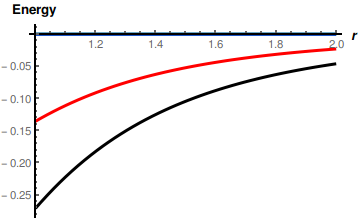}
\end{subfigure}
\caption{\fcolorbox{red}{red}{\rule{0pt}{2.5pt}\rule{0pt}{2.5pt}} $\rho$, 
 \fcolorbox{black}{black}{\rule{0pt}{2.5pt}\rule{0pt}{2.5pt}} $\rho+p_r$, 
 \fcolorbox{green}{green}{\rule{0pt}{2.5pt}\rule{0pt}{2.5pt}} $\rho+p_t$, 
 \fcolorbox{blue}{blue}{\rule{0pt}{2.5pt}\rule{0pt}{2.5pt}} $\rho+p_r+2p_t$, The plot depicts $\rho$, $\rho+p_r$, $\rho+p_t$ and $\rho+p_r+2p_t$. It shows that NEC, WEC, SEC is violated throughout the space-time.
 }
\label{FIG.1} 
\end{figure}

For the exponential metric, the explicit forms of the non-zero Einstein tensor components are,
\begin{equation}
G^r_r=-G^t_t=-G^{\theta}_{\theta}=-G^{\phi}_{\phi}=-\frac{m^2 e^{-\frac{2m}{r}}}{r^4}.
\end{equation}
In the units of $c=1$, $G=1$, it leads to,
\begin{equation}
\rho(r)=p_r(r)=-p_t(r)=-\frac{m^2 e^{-\frac{2m}{r}}}{r^4}.
\end{equation}
Where $\rho$, $p_r$ and $p_t$ stand for energy density, radial and tangential pressure respectively. From the above relation we can see that,
\begin{equation}
\rho+p_r=-\frac{2 m^2 e^{-\frac{2m}{r}}}{r^4}<0,
\end{equation}
\begin{equation}
\rho + p_t=0
\end{equation}
and 
\begin{equation}
\rho + p_r+ 2 p_t=0
\end{equation}
At the throat, the above takes the value,
\begin{equation}
 \rho+p_r= -\frac{2}{(me)^2}<0;\rho + p_t=0;\rho + p_r+ 2 p_t=0
\end{equation}
and
\begin{equation}
\rho =-\frac{1}{(me)^2}<0.
\end{equation}
In terms of the principal pessures, the energy conditions are given as,
\\ Null Energy Condition(NEC): $\rho+p_r\geq0$, $\rho+p_t\geq 0$
\\ Weak Energy Condition(WEC): $\rho\geq0$, $\rho+p_r\geq0$, $\rho+p_t\geq 0$
\\ Strong Energy Condition(SEC): $\rho\geq0$, $\rho+p_t\geq0$, $\rho+p_r+2 p_t\geq0$
\\ It is seen that the exponential wormhole metric partially violates all the energy conditions, specially at or near the throat. Which can be more clearly visible from the FIG(\ref{FIG.1}).
\subsection{Flare-out condition}
The flare-out condition is more understandable through the embedding geometry. The embedded spacetime at $t=$constant and $\theta=\frac{\pi}{2}$ for the exponential wormhole metric is given by,
\begin{equation}\label{(embed)}
ds^2_e= e^{\frac{2m}{r}}[dr^2+ r^2 d\phi^2]
\end{equation}
In three dimensional Euclidean space the embedded surface has equation $z=z(r)$, so that the metric of the surface can be written as,
\begin{equation}\label{(z)}
ds^2_e=[1+(\frac{dz}{dr})^2]dr^2+e^{\frac{2m}{r}}r^2 d\phi^2
\end{equation}
Comparing the relations Eq.(\ref{(embed)}) and Eq.(\ref{(z)}), we get
\begin{equation}
\frac{dz}{dr}=\pm(e^{\frac{2m}{r}}-1)^{\frac{1}{2}}.
\end{equation}
Here, we observe that $\frac{dz}{dr}\to 0$ as $r\to \infty$. Which implies that the space is asymptotically flat \cite{kim2013flare}. Now, the flare-out condition is given by the minimality of the wormhole throat as,
\begin{equation}
\frac{d}{dz}\left(\frac{dr}{dz}\right)=\frac{m e^{\frac{2m}{r}}}{r^2(e^{\frac{2m}{r}}-1)^2}>0
\end{equation}
i.e., $m>0$. Again, for the exponential wormhole metric, surface tension $\tau$ is given as \cite{kim2013flare, morris1988wormholes},
\begin{equation}
\tau=\frac{m^2 e^{-\frac{2m}{r}}}{r^4}
\end{equation}
Usually, the exoticity function $\zeta$ is used for the flare-out condition, which is given as,
\begin{equation}
\zeta=\frac{\tau-\rho}{|\rho|}>0.
\end{equation}
For the exponential wormhole metric, the value of $\zeta$ comes out as,
\begin{equation}
\zeta=\frac{\tau-\rho}{|\rho|}=2>0
\end{equation}
So, for the exponential wormhole metric $(\tau-\rho)>0$ everywhere, i.e. the metric obeys flare-out conditions everywhere. It was assumed that the wormhole should have a large surface tension compared to the energy density to continue the geometry. This condition seems to be physically reasonable. This violates the Weak Energy Condition(WEC)or averaged Weak Energy Condition to minimize the use of exotic matter. 
\subsection{Curvature tensor}
The non-zero components of Riemann tensor, Rici tensor, Ricci curvature scalar, Kretschmann scalar and other related scalars for the exponential metric are,
\begin{equation}
R^{tr}_{tr}=-2R^{t\theta}_{t\theta}=-2R^{t\phi}_{t\phi}=\frac{2me^{-\frac{2m}{r}}(r-m)}{r^4},
\end{equation}
\begin{equation}
R^{r\theta}_{r\theta}=R^{r\phi}_{r\phi}=-\frac{m}{r^3}e^{-\frac{2m}{r}},
\end{equation}
\begin{equation}
R^{\theta \phi}_{\theta \phi}=\frac{m(2r-m)}{r^4}e^{-\frac{2m}{r}},
\end{equation}
\begin{equation}
R^a_b=-\frac{2m^2}{r^4}e^{-\frac{2m}{r}} diag(0,1,0,0)^a_b,
\end{equation}
\begin{equation}
R=-\frac{2m^2}{r^4}e^{-\frac{2m}{r}},
\end{equation}
\begin{equation}
R_{abcd}R^{abcd}=\frac{4m^2(12r^2-16mr+7m^2)}{r^8}e^{-\frac{2m}{r}},
\end{equation}
\begin{equation}
C_{abcd}C^{abcd}=\frac{16m^2(3r-2m)}{3r^8}e^{-\frac{2m}{r}},
\end{equation}
\begin{equation}
R_{ab}R^{ab}=\frac{4m^4}{r^8}e^{-\frac{2m}{r}}
\end{equation}
The non zero Electric parts of the Weyl tensors are,
\begin{equation}
E_{\theta \theta}=E_{\phi \phi}=-2E_{rr}=\frac{2m(r-m)}{3r^4}e^{-\frac{2m}{r}},
\end{equation}
\begin{equation}
E_{tt}=-\frac{m(m+2r)}{3r^4}e^{-\frac{6m}{r}},
\end{equation}
\begin{equation}
E_{ab}E^{ab}=\frac{m^2(-4mr(2+7r^4)+m^2(4+17r^4)+4(r^2+5r^6)+4(m-r)^2 \csc^4\theta}{9r^{12}}e^{-\frac{8m}{r}}
\end{equation}
All these components are finite and they donot diverge at $r=0$ and $r=m$, they decreases to zero both as $r\to \infty$ and as $r\to 0$. So we can say that the exponential wormhole metric doesnot contain any kind of Weyl and Oscillating Ricci singularity[ref].
\par
\begin{figure}[H]
\begin{subfigure}[b]{0.4\textwidth}
\includegraphics[width=\textwidth]{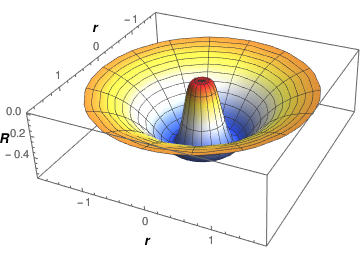}
\caption{Plot of $R$}
\end{subfigure}
\hfill
\begin{subfigure}[b]{0.4\textwidth}
\includegraphics[width=\textwidth]{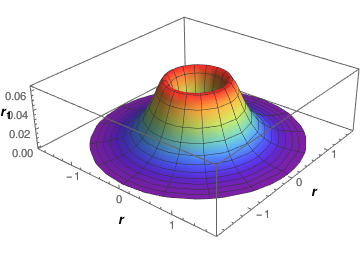} 
\caption{Plot of $r_1$} 
\end{subfigure}
\hfill
\begin{subfigure}[b]{0.4\textwidth}
\includegraphics[width=\textwidth]{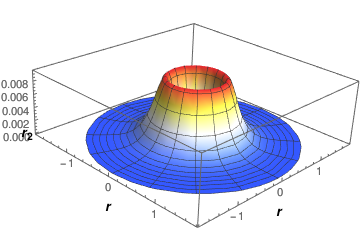} 
\caption{Plot of $r_2$} 
\end{subfigure}
\hfill
\begin{subfigure}[b]{0.4\textwidth}
\includegraphics[width=\textwidth]{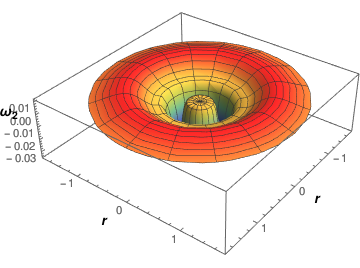} 
\caption{Plot of $\omega_2$} 
\end{subfigure}
\caption{The plots depict the four curvature invariants of the exponential wormhole metric. Here we have considered $m=1$, i.e. the throat is at $r=1$.}
\label{FIG.2}
\end{figure}

 The four curvature invariants viz. Ricci scalar, the first two Ricci invariants and the real component of the Weyl invariant are \cite{mattingly2020curvature, carminati1991algebraic},
\begin{equation}
R= -\frac{2m^2}{r^4}e^{-\frac{2m}{r}},
\end{equation}
\begin{equation}
r_1=\frac{1}{4}S^b_a S^a_b =\frac{3m^4}{4r^8}e^{-\frac{4m}{r}},
\end{equation}
\begin{equation}
r_2=-\frac{1}{8}S^b_a S^a_c S^c_b=\frac{3m^6}{8r^{12}}e^{-\frac{6m}{r}}
\end{equation}
and
\begin{equation}
\omega_2=-\frac{1}{8} \bar{C}_{abcd} \bar{C}^{abef} \bar{C}^{cd}_{ef}=-\frac{32m^3(2m-3r)^3}{9r^{12}}e^{-\frac{2m}{r}}.
\end{equation}
Where $S_{ab}=R_{ab}-\frac{1}{4}g_{ab}R$. When the curvature invariants are plotted FIG(\ref{FIG.2}), it is seen that they are all nonzero and depend only on the radial coordinate $r$ which indicates spherical symmetry. Again they are finite at $r=m$(at the throat) and decay to zero as $r\to \infty$. $R$ and $\omega_2$ have a minima near the throat whereas $r_1$ and $r_2$ have a maxima. These plots are finite everywhere indicating the absence of horizon. So we can conclude that the exponential wormhole metric represents a traversable wormhole.
\subsection{Ricci convergence}
Any Lorentzian spacetime is said to fulfil the timelike, null and spacelike Ricci convergence condition if for all timelike, null or spacelike vectors $t^a$ one has \cite{boonserm2018exponential, visser1995lorentzian},
\begin{equation}
R_{ab}t^at^b \geq 0
\end{equation}
Now, for the exponential wormhole metric, we can see that,
\begin{equation}
R_{ab}=-\frac{2m^2}{r^4}diag(0,1,0,0)_{ab}.
\end{equation}
So the Ricci convergence condition leads to,
\begin{equation}
R_{ab}t^at^b=-\frac{2m^2}{r^4}(t^r)^2 \leq 0.
\end{equation}
So, the exponential wormhole metric violates the null Ricci convergence condition for all timelike, null and spacelike vectors. Again we can show that,
\begin{equation}
R_{ab}=-\frac{2m^2}{r^4}diag(0,1,0,0)_{ab}=-\frac{1}{2}\nabla_a(\frac{2m}{r})\nabla_b(\frac{2m}{r})=-\frac{1}{2}\nabla_a\Phi \nabla_b\Phi.
\end{equation}
and
\begin{equation}
G_{ab}=-\frac{1}{2}[\nabla_a\Phi \nabla_b\Phi-\frac{1}{2}g_{ab}(g^{cd}\nabla_c \Phi \nabla_d \Phi)]
\end{equation}
i.e. Einstein equation for a negative kinetic energy massless scalar field, a ghost or phantom field. The contracted Bianchi identity $G^{ab}_{;b}$ gives the scalar field equation of motion $(g^{ab} \nabla_a \nabla_b)\Phi=0$, which indicates that the exponential wormhole metric represents a traversable wormhole metric \cite{morris1988wormholes, visser1995lorentzian}.
\section{Exponential wormhole metric in $f(R)$ Gravity}
The gravitational action for $f(R)$ gravity can be defined as,
\begin{equation}\label{(L)}
S=\frac{1}{2k} \int [f(R)+L_m]\sqrt{-g}d^4x,
\end{equation}
where $k=8\pi G$, $L_m$ and $g$ stand for the matter Lagrangian density and the determinant of the metric $g_{\mu \nu}$ respectively. Here, for simplicity, we will consider $k$ as unity. Now varying the Eq.(\ref{(L)}) with respect to the metric $g_{\mu\nu}$ gives the field equations as,
\begin{equation}\label{(v)}
FR_{\mu\nu}-\frac{1}{2}fg_{\mu\nu}-\nabla_{\mu}\nabla_{\nu}F+ \Box F g_{\mu\nu}=T^m_{\mu\nu},
\end{equation}
where $R_{\mu\nu}$ represents Ricci tensor and $F=\frac{df}{dR}$. Now we can consider the contraction of Eq.(\ref{(v)}) to obtain the relation,
\begin{equation}\label{(c)}
FR-2f+3 \Box F=T.
\end{equation}
Where $R=g^{\mu\nu}R_{\mu\nu}$ and $T=g^{\mu\nu}T_{\mu\nu}$ represent Ricci scalar and trace of stress energy tensor respectively. Combining the Eqs.(\ref{(v)}) and Eq.(\ref{(c)}), the effective field equations are calculated as,
\begin{equation}
G_{\mu\nu}\equiv R_{\mu\nu}-\frac{1}{2}Rg_{\mu\nu}=T^{eff}_{\mu\nu},\;\;\;\;with\;\;\;\;T^{eff}_{\mu\nu}=T^c_{\mu\nu}+\frac{T^m_{\mu\nu}}{F},
\end{equation}
where
\begin{equation}
T^c_{\mu\nu}=\frac{1}{F}[\nabla_{\mu} \nabla_{\nu} F-\frac{1}{4}(FR+\Box F+T)].
\end{equation}
The energy momentum tensor for the matter source of the wormholes is $T_{\mu\nu}=\frac{\partial L_m}{\partial g^{\mu\nu}}$, which is defined as,
\begin{equation}
T_{\mu\nu}=(\rho+p_t)u_{\mu} u_{\nu}-p_t g_{\mu\nu}+(p_r-p_t)X_{\mu} X_{\nu},
\end{equation}
such that
\begin{equation}
u^{\mu}u_{\nu}=-1 \;\;\;\;and\;\;\;\;X^{\mu}X_{\nu}=1,
\end{equation}
where $u_{\mu}$ is the four velocity and $X_{\mu}$ is the unit space-like vector. Again $\rho$, $p_r$ and $p_t$ are energy density, radial pressure and tangential pressure respectively.  Now Einstein's field equation for the metric Eq.(\ref{(expo)}) in $f(R)$ gravity can be solved as,
\begin{equation}\label{(f1)}
\rho=-\frac{e^{-\frac{2m}{r}}(e^{\frac{2m}{r}}Hr^4+m^2 F(r)+mr^2F^{\prime}(r))}{r^4},
\end{equation}
\begin{equation}\label{(f2)}
p_r=\frac{e^{-\frac{2m}{r}}(e^{\frac{2m}{r}}Hr^4-m^2F(r)+mr^2 F^{\prime}(r)+r^4 F^{\prime \prime}(r))}{r^4}
\end{equation}
and
\begin{equation}\label{(f3)}
p_t=\frac{e^{\frac{2m}{r}}(e^{\frac{2m}{r}}Hr^4+m^2F(r)-mr^2F^{\prime}(r)+r^3F^{\prime}(r))}{r^4}
\end{equation}
where, $H=\frac{1}{4}(FR+\Box F+T)$ and $F^{\prime}=\frac{dF(r)}{dr}$ and $F^{\prime \prime}=\frac{d^2 F(r)}{dr^2}$.
\par Bronnikov and Starobinsky \cite{bronnikov2007no} considered the stability condition for wormhole geometry which is free from ghosts. They showed that no realistic wormhole can be constructed in scalar-tensor models for a positive scalar function. In $f(R)$ gravity, the non-existence of wormhole could be disobeyed if $\frac{df}{dR}=F(R)$ is negative \cite{bronnikov201016}. In agreement with classical General Theory of Relativity, the violation of Null Energy Condition(NEC) denoted as $\rho+p_r \geq 0$, $\rho+p_t \geq 0$ and Weak Energy Condition (WEC) denoted as $\rho \geq 0$, $\rho+p_r \geq 0$, $\rho+p_t \geq 0$ are mainly due to the presence of exotic matter. The wormhole throat mainly doesnot respect the NEC \cite{hochberg1998dynamic, hochberg1998null}. In order to check the the necessary NEC, we simplify our calculations for $\rho+p_r$ and $\rho+p_t$.  For the exponential wormhole metric, to obey the NEC,
\begin{equation}
\rho+p_r=\frac{e^{-\frac{2m}{r}}(-2m^2 F(r)+r^4 F^{\prime \prime}(r))}{r^4}
\end{equation} 
and
\begin{equation}
\rho+p_t= \frac{e^{-\frac{2m}{r}}(-2m+r)F^{\prime}(r)}{r^2}
\end{equation}
should be positive at the throat. In the next section, we will discuss exponential wormhole solutions under the influence of four different viable $f(R)$ gravity models.
\subsection{The Exponential Gravity Model}
The exponential gravity model was introduced and investigated by Cognola \cite{cognola2008class}. This model can describe the inflation of early universe and accelerated expansion of the current universe. The exponential model is defined as,
\begin{equation}
f(R)= R-\mu R_0[1-e^{-\frac{R}{R_0}}]
\end{equation}
where $\mu$ and $R_0$ are arbitrary constants. Now the equations Eq.(\ref{(f1)}), Eq.(\ref{(f2)}) and Eq.(\ref{(f3)}) reduces to,
\begin{equation}
\rho= \frac{e^{-\frac{4m}{r}}}{r^8 R_0}[-e^{\frac{2m}{r}}r^4(m^2+e^{\frac{2m}{r}}Hr^4)R_0+e^{\frac{2m^2 e^{-\frac{2m}{r}}}{r^4 R_0}}m^2(4m(m-2r)+e^{\frac{2m}{r}}r^4 R_0)\mu],
\end{equation}
\begin{align}
\begin{split}
p_r={}&\frac{e^{-\frac{6m}{r}}}{r^{12} R_0^2}[-16 e^{\frac{2m^2 e^{-\frac{2m}{r}}}{r^4 R_0}}m^4(m-2r)^2\mu -r^4 R_0[e^{\frac{4m}{r}}r^4[-m^2+Hr^4 e^{\frac{2m}{r}}]R_0+ e^{\frac{2m(r^3+\frac{m e^{-\frac{2m}{r}}}{R_0})}{r^4}}m^2\\
&[-12m^2+48mr-40r^2+r^4 R_0 e^{\frac{2m}{r}}]\mu]]
\end{split}
\end{align}
and
\begin{equation}
p_t= \frac{e^{-\frac{4m}{r}}}{r^8 R_0}[e^{\frac{2m}{r}}r^4(m^2+e^{\frac{2m}{r}}Hr^4)R_0+e^{\frac{2m^2 e^{-\frac{2m}{r}}}{r^4 R_0}}m^2(4m^2-12mr+8r^2-e^{\frac{2m}{r}}r^4 R_0)\mu]
\end{equation}
Here we use $m=1$ and evaluate the graphical behaviour of $\rho$, $\rho+p_r$, $\rho+p_t$, $\rho+p_r+2p_t$ and $F=\frac{df}{dR}$
\begin{figure}[H]
\begin{subfigure}[b]{0.4\textwidth}
\includegraphics[width=\textwidth]{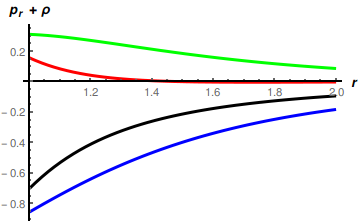}
\caption{\fcolorbox{red}{red}{\rule{0pt}{2.5pt}\rule{0pt}{2.5pt}} $\mu=+ve,R_0=+ve$, 
 \fcolorbox{black}{black}{\rule{0pt}{2.5pt}\rule{0pt}{2.5pt}} $\mu=-ve,R_0=+ve$, 
 \fcolorbox{green}{green}{\rule{0pt}{2.5pt}\rule{0pt}{2.5pt}} $\mu=+ve,R_0=-ve$, 
 \fcolorbox{blue}{blue}{\rule{0pt}{2.5pt}\rule{0pt}{2.5pt}} $\mu=-ve,R_0=-ve$}
\end{subfigure}
\hfill
\begin{subfigure}[b]{0.4\textwidth}
\includegraphics[width=\textwidth]{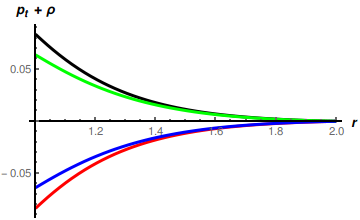} 
\caption{\fcolorbox{red}{red}{\rule{0pt}{2.5pt}\rule{0pt}{2.5pt}} $\mu=+ve,R_0=+ve$, 
 \fcolorbox{black}{black}{\rule{0pt}{2.5pt}\rule{0pt}{2.5pt}} $\mu=-ve,R_0=+ve$, 
 \fcolorbox{green}{green}{\rule{0pt}{2.5pt}\rule{0pt}{2.5pt}} $\mu=+ve,R_0=-ve$, 
 \fcolorbox{blue}{blue}{\rule{0pt}{2.5pt}\rule{0pt}{2.5pt}} $\mu=-ve,R_0=-ve$} 
\end{subfigure}
\hfill
\begin{subfigure}[b]{0.4\textwidth}
\includegraphics[width=\textwidth]{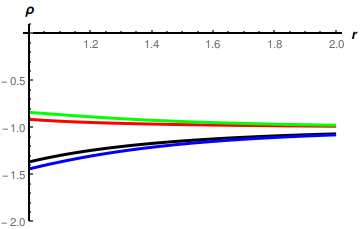} 
\caption{\fcolorbox{red}{red}{\rule{0pt}{2.5pt}\rule{0pt}{2.5pt}} $\mu=+ve,R_0=+ve$, 
 \fcolorbox{black}{black}{\rule{0pt}{2.5pt}\rule{0pt}{2.5pt}} $\mu=-ve,R_0=+ve$, 
 \fcolorbox{green}{green}{\rule{0pt}{2.5pt}\rule{0pt}{2.5pt}} $\mu=+ve,R_0=-ve$, 
 \fcolorbox{blue}{blue}{\rule{0pt}{2.5pt}\rule{0pt}{2.5pt}} $\mu=-ve,R_0=-ve$} 
\end{subfigure}
\hfill
\begin{subfigure}[b]{0.4\textwidth}
\includegraphics[width=\textwidth]{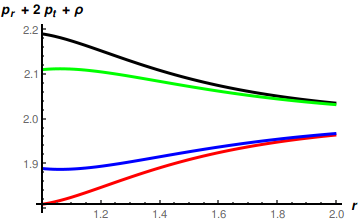} 
\caption{\fcolorbox{red}{red}{\rule{0pt}{2.5pt}\rule{0pt}{2.5pt}} $\mu=+ve,R_0=+ve$, 
 \fcolorbox{black}{black}{\rule{0pt}{2.5pt}\rule{0pt}{2.5pt}} $\mu=-ve,R_0=+ve$, 
 \fcolorbox{green}{green}{\rule{0pt}{2.5pt}\rule{0pt}{2.5pt}} $\mu=+ve,R_0=-ve$, 
 \fcolorbox{blue}{blue}{\rule{0pt}{2.5pt}\rule{0pt}{2.5pt}} $\mu=-ve,R_0=-ve$} 
\end{subfigure}
\hfill
\begin{subfigure}[b]{0.4\textwidth}
\includegraphics[width=\textwidth]{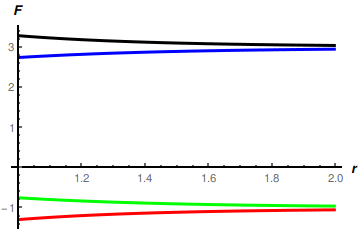} 
\caption{\fcolorbox{red}{red}{\rule{0pt}{2.5pt}\rule{0pt}{2.5pt}} $\mu=+ve,R_0=+ve$, 
 \fcolorbox{black}{black}{\rule{0pt}{2.5pt}\rule{0pt}{2.5pt}} $\mu=-ve,R_0=+ve$, 
 \fcolorbox{green}{green}{\rule{0pt}{2.5pt}\rule{0pt}{2.5pt}} $\mu=+ve,R_0=-ve$, 
 \fcolorbox{blue}{blue}{\rule{0pt}{2.5pt}\rule{0pt}{2.5pt}} $\mu=-ve,R_0=-ve$} 
\end{subfigure}
\caption{The plots depict $\rho+p_r$((a)), $\rho+p_t$((b)), $\rho$((c)), $\rho+p_r+2p_t$((d)) and $F=\frac{df}{dR}$((e)). It shows that NEC is respected throughout the space-time for some specific combinations of $\mu$ and $R_0$ while there is a partial violation of WEC and SEC.}
\label{FIG.3}
\end{figure}
\par From the graph we observe that,
\begin{itemize}
\item If $\mu=+ve$, $R_0=-ve$, then $\rho+p_r \geq 0$
\item If $\mu=+ve$, $R_0=-ve$ or $\mu = -ve$, $R_0= +ve$, then $\rho+p_t \geq 0$
\item $\rho \leq 0$ for all combinations of values of $\mu$ and $R_0$
\item $\rho+p_r+2 p_t \geq 0$ for all combinations of value of $\mu$ and $R_0$.
\item $\frac{df}{dR}>0$ for $\mu=-ve, R_0=+ve$ or $\mu=-ve, R_0=-ve$ and $\frac{df}{dR}<0$ for $\mu=+ve, R_0=-ve$ or $\mu=+ve,R_0=+ve$.
\end{itemize}
So we can conclude that if $\mu = +ve$, $R_0= -ve$, the necessary NEC is respected throughout the wormhole geometry and $F=\frac{df}{dR}$<0, but WEC and SEC is partially violated. So for this combination of $\mu$ and $R_0$, we get wormhole solution which violates the non-existence theorem with the presence of negligible amount of exotic matter. Whereas in the case of General Relativity, NEC is violated by exponetial wormhole metric.
\subsection{Starobinsky $f(R)$ gravity model}
\begin{figure}[H]
\begin{subfigure}[b]{0.4\textwidth}
\includegraphics[width=\textwidth]{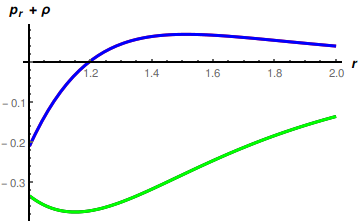}
\caption{\fcolorbox{green}{green}{\rule{0pt}{2.5pt}\rule{0pt}{2.5pt}} $a=-ve,R_0=+ve\;\;or\;\;a=+ve,R_0=-ve$, 
 \fcolorbox{blue}{blue}{\rule{0pt}{2.5pt}\rule{0pt}{2.5pt}} $a=+ve,R_0=+ve\;\;or\;\;a=-ve,R_0=-ve$}
\end{subfigure}
\hfill
\begin{subfigure}[b]{0.4\textwidth}
\includegraphics[width=\textwidth]{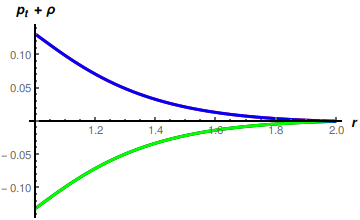} 
\caption{\fcolorbox{green}{green}{\rule{0pt}{2.5pt}\rule{0pt}{2.5pt}} $a=-ve,R_0=+ve\;\;or\;\;a=+ve,R_0=-ve$, 
 \fcolorbox{blue}{blue}{\rule{0pt}{2.5pt}\rule{0pt}{2.5pt}} $a=+ve,R_0=+ve\;\;or\;\;a=-ve,R_0=-ve$} 
\end{subfigure}
\hfill
\begin{subfigure}[b]{0.4\textwidth}
\includegraphics[width=\textwidth]{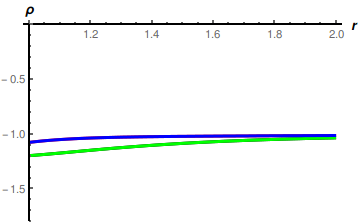} 
\caption{\fcolorbox{green}{green}{\rule{0pt}{2.5pt}\rule{0pt}{2.5pt}} $a=-ve,R_0=+ve\;\;or\;\;a=+ve,R_0=-ve$, 
 \fcolorbox{blue}{blue}{\rule{0pt}{2.5pt}\rule{0pt}{2.5pt}} $a=+ve,R_0=+ve\;\;or\;\;a=-ve,R_0=-ve$} 
\end{subfigure}
\hfill
\begin{subfigure}[b]{0.4\textwidth}
\includegraphics[width=\textwidth]{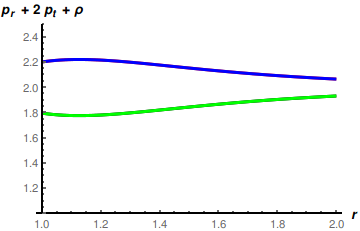} 
\caption{\fcolorbox{green}{green}{\rule{0pt}{2.5pt}\rule{0pt}{2.5pt}} $a=-ve,R_0=+ve\;\;or\;\;a=+ve,R_0=-ve$, 
 \fcolorbox{blue}{blue}{\rule{0pt}{2.5pt}\rule{0pt}{2.5pt}} $a=+ve,R_0=+ve\;\;or\;\;a=-ve,R_0=-ve$} 
\end{subfigure}
\hfill
\begin{subfigure}[b]{0.4\textwidth}
\includegraphics[width=\textwidth]{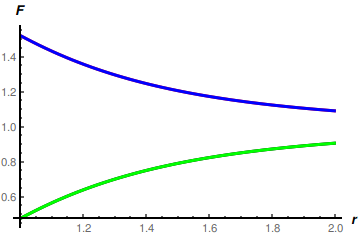} 
\caption{\fcolorbox{green}{green}{\rule{0pt}{2.5pt}\rule{0pt}{2.5pt}} $a=-ve,R_0=+ve\;\;or\;\;a=+ve,R_0=-ve$, 
 \fcolorbox{blue}{blue}{\rule{0pt}{2.5pt}\rule{0pt}{2.5pt}} $a=+ve,R_0=+ve\;\;or\;\;a=-ve,R_0=-ve$} 
\end{subfigure}
\caption{The plots depict $\rho+p_r$((a)), $\rho+p_t$((b)), $\rho$((c)), $\rho+p_r+2p_t$((d)) and $F=\frac{df}{dR}$((e)) for $l=+ve$. It shows that NEC is respected throughout the space-time for some specific combination of $a$ and $R_0$ while there is a partial violation of WEC and SEC.}
\label{FIG.4}
\end{figure}

\begin{figure}[H]
\begin{subfigure}[b]{0.4\textwidth}
\includegraphics[width=\textwidth]{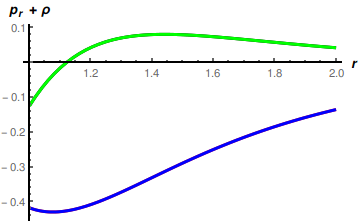}
\caption{\fcolorbox{green}{green}{\rule{0pt}{2.5pt}\rule{0pt}{2.5pt}} $a=-ve,R_0=+ve\;\;or\;\;a=+ve,R_0=-ve$, 
 \fcolorbox{blue}{blue}{\rule{0pt}{2.5pt}\rule{0pt}{2.5pt}} $a=+ve,R_0=+ve\;\;or\;\;a=-ve,R_0=-ve$}
\end{subfigure}
\hfill
\begin{subfigure}[b]{0.4\textwidth}
\includegraphics[width=\textwidth]{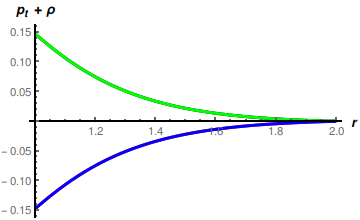} 
\caption{\fcolorbox{green}{green}{\rule{0pt}{2.5pt}\rule{0pt}{2.5pt}} $a=-ve,R_0=+ve\;\;or\;\;a=+ve,R_0=-ve$, 
 \fcolorbox{blue}{blue}{\rule{0pt}{2.5pt}\rule{0pt}{2.5pt}} $a=+ve,R_0=+ve\;\;or\;\;a=-ve,R_0=-ve$} 
\end{subfigure}
\hfill
\begin{subfigure}[b]{0.4\textwidth}
\includegraphics[width=\textwidth]{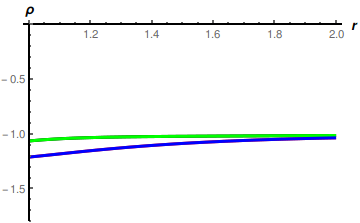} 
\caption{\fcolorbox{green}{green}{\rule{0pt}{2.5pt}\rule{0pt}{2.5pt}} $a=-ve,R_0=+ve\;\;or\;\;a=+ve,R_0=-ve$, 
 \fcolorbox{blue}{blue}{\rule{0pt}{2.5pt}\rule{0pt}{2.5pt}} $a=+ve,R_0=+ve\;\;or\;\;a=-ve,R_0=-ve$} 
\end{subfigure}
\hfill
\begin{subfigure}[b]{0.4\textwidth}
\includegraphics[width=\textwidth]{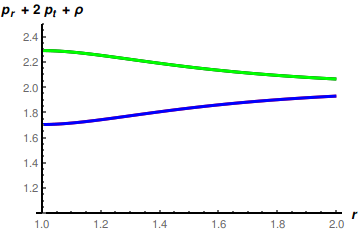} 
\caption{\fcolorbox{green}{green}{\rule{0pt}{2.5pt}\rule{0pt}{2.5pt}} $a=-ve,R_0=+ve\;\;or\;\;a=+ve,R_0=-ve$, 
 \fcolorbox{blue}{blue}{\rule{0pt}{2.5pt}\rule{0pt}{2.5pt}} $a=+ve,R_0=+ve\;\;or\;\;a=-ve,R_0=-ve$} 
\end{subfigure}
\hfill
\begin{subfigure}[b]{0.4\textwidth}
\includegraphics[width=\textwidth]{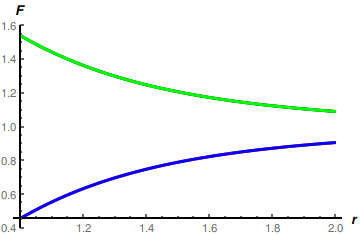} 
\caption{\fcolorbox{green}{green}{\rule{0pt}{2.5pt}\rule{0pt}{2.5pt}} $a=-ve,R_0=+ve\;\;or\;\;a=+ve,R_0=-ve$, 
 \fcolorbox{blue}{blue}{\rule{0pt}{2.5pt}\rule{0pt}{2.5pt}} $a=+ve,R_0=+ve\;\;or\;\;a=-ve,R_0=-ve$} 
\end{subfigure}
\caption{The plots depict $\rho+p_r$((a)), $\rho+p_t$((b)), $\rho$((c)), $\rho+p_r+2p_t$((d)) and $F=\frac{df}{dR}$((e)) for $l=-ve$. It shows that NEC is respected throughout the space-time for some specific combination of $a$ and $R_0$ while there is a partial violation of WEC and SEC.}
\label{FIG.5}
\end{figure}

This model was proposed by Satrobinsky \cite{starobinsky2007disappearing} which is one of the most recognized $f(R)$ gravity model. It is consistent with cosmological conditions and satisfies solar system and laboratory tests. Starobinsky model is given as,
\begin{equation}
f(R)=R+a R_0[(1+\frac{R^2}{R_0^2})^{-l}-1],
\end{equation}
where $a$, $R_0$ and $l$ are free parameters. The field equations Eq.(\ref{(f1)}), Eq.(\ref{(f2)}) and Eq.(\ref{(f3)}) of the exponential wormhole metric in $f(R)$ gravity model reduce as,
\begin{align}
\begin{split}
\rho={}&-H+m^2 [-\frac{e^{-\frac{2m}{r}}}{r^4}+\\
& \frac{4alm(1+\frac{4m^4 e^{-\frac{4m}{r}}}{R_0^2 r^8})^{-l}R_0(4m^4(m+4lm-4(r+2lr))+e^{\frac{4m}{r}}(-3m+4r)r^8R_0^2)}{(4m^4+ R_0^2 r^8 e^{\frac{4m}{r}})^2}],
\end{split}
\end{align}
\begin{align}
\begin{split}
p_r={}& \frac{1}{r^4(4m^4+r^8 R_0^2 e^{\frac{4m}{r}})^3} e^{-\frac{2m}{r}}[1+\frac{4m^4 e^{-\frac{4m}{r}}}{r^8 R_0^2}]^{-l}[64a e^{\frac{2m}{r}}lm^{12}r^4 R_0+768 e^{\frac{2m}{r}}l^2 m^{12}r^4 R_0+\\
&1024a e^{\frac{2m}{r}}l^3m^{12}r^4 R_0-512ae^{\frac{2m}{r}}lm^{11}r^5R_0-3072ae^{\frac{2m}{r}}l^2 m^{11}r^5R_0-4096ae^{\frac{2m}{r}}l^3 m^{11}r^5R_0\\
&+768ae^{\frac{2m}{r}}lm^{10}r^6R_0+3584ae^{\frac{2m}{r}}l^2m^{10}r^6R_0+4096ae^{\frac{2m}{r}}l^3m^{10}r^6R_0-416ae^{\frac{2m}{r}}lm^8 r^{12}R_0^3\\
&-576ae^{\frac{6m}{r}}l^2m^8r^{12}R_0^3+1536ae^{\frac{6m}{r}}lm^7r^{13}R_0^3+2304ae^{\frac{6m}{r}}l^2m^7r^{13}R_0^3-1536ae^{\frac{2m}{r}}lm^6r^{14}R_0^3\\
&-2176ae^{\frac{6m}{r}}l^2m^6r^{14}R_0^3+20ae^{\frac{10m}{r}}lm^4r^{20}R_0^5-96ae^{\frac{10m}{r}}km^3r^{21}R_0^5+80ae^{\frac{10m}{r}}lm^2r^{22}R_0^5\\
&-(m^2-Hr^4e^{\frac{2m}{r}})(1+\frac{4m^4e^{-\frac{4m}{r}}}{r^8R_0^2})^l(4m^4+r^8R_0^2 e^{\frac{4m}{r}})^3]
\end{split}
\end{align}
and
\begin{align}
\begin{split}
p_t={}& H+m^2[\frac{e^{-\frac{2m}{r}}}{r^4}+\\
& \frac{1}{(4m^4+r^8 R_0^2e^{\frac{4m}{r}})^2}[4al(1+\frac{4m^4 e^{-\frac{4m}{r}}}{r^8R_0^2})^{-l}R_0(4m^4((3+4l)m^2+4(1+2l)r^2-6m(r+\\
&2lr))-e^{\frac{4m}{r}}r^8(m^2-6mr+4r^2)R_0^2)]]
\end{split}
\end{align}
After plotting the graphs FIG(\ref{FIG.4}) and FIG(\ref{FIG.5}) of $\rho$, $\rho+p_r$, $\rho+p_t$, $\rho+p_r+2p_t$ and $F=\frac{df}{dR}$, we get the following analysis,
\par If $l=+ve$
\begin{itemize}
\item If $a=+ve$, $R_0=+ve$ or $a=-ve$, $R_0=-ve$, then $\rho+p_r \geq 0$ (for $r>1.2$).
\item If $a=+ve$, $R_0=+ve$ or $a=-ve$, $R_0=-ve$, then $\rho+p_t \geq 0$.
\item $\rho \leq 0$ for all combinations of $a$ and $R_0$.
\item $\rho+p_r+2p_t \geq 0$ for all combinations of $a$ and $R_0$.
\item $\frac{df}{dR}>0$ for all combinations of $a$ and $R_0$.
\end{itemize}
\par If $l=-ve$
\begin{itemize}
\item If $a=-ve$, $R_0=+ve$ or $a=+ve$, $R_0=-ve$, then $\rho+p_r \geq 0$ (for $r>1.13$).
\item If $a=-ve$, $R_0=+ve$ or $a=+ve$, $R_0=-ve$, then $\rho+p_r \geq 0$.
\item $\rho \leq 0$ for all combinations of $a$ and $R_0$.
\item $\rho+p_r+2p_t \geq 0$ for all combinations of $a$ and $R_0$.
\item $\frac{df}{dR}>0$ for all combinations of $a$ and $R_0$.
\end{itemize}
So we conclude that, if $l=+ve$, $a=+ve$, $R_0=+ve$ or $l=+ve$, $a=-ve$, $R_0=-ve$ (for $r>1.2$) and $l=-ve$, $a=-ve$, $R_0=+ve$ or $l=-ve$, $a=+ve$, $R_0=-ve$ (for $r>1.33$), then NEC is respected throughout the geometry. But NEC is partially violated at the throat, that is to show the feasible traversable wormhole structure which have small amount of exotic matter at the throat of the wormhole. Again $F=\frac{df}{dR}>0$ represents the non-spherically symmetric wormhole solution.
\subsection{Tsujikawa $f(R)$ gravity model}
This model was represented by Tsujikawa \cite{tsujikawa2008observational} and it is defined as,
\begin{equation}
f(R)=R-\mu R_0 \tanh[\frac{R}{R_0}],
\end{equation}
where $\mu$ and $R_0$ are arbitrary constants. The field equations Eq.(\ref{(f1)}), Eq.(\ref{(f2)}) and Eq.(\ref{(f3)}) now reduce to,
\begin{align}
\begin{split}
\rho={}& \frac{e^{-\frac{4m}{r}}}{r^8 R_0}[-e^{\frac{2m}{r}}r^4[m^2+Hr^4 e^{\frac{2m}{r}}]R_0+m^2 \mu  \sech[\frac{2m^2 e^{-\frac{2m}{r}}}{r^4 R_0}]^2[r^4 R_0 e^{\frac{2m}{r}}-\\
&8m(m-2r)\tanh[\frac{2m^2 e^{-\frac{2m}{r}}}{r^4 R_0}]]],
\end{split}
\end{align}
\begin{align}
\begin{split}
p_r={}& \frac{e^{-\frac{6m}{r}}}{r^{12}R_0^2}[e^{\frac{4m}{r}}r^8[-m^2+ e^{\frac{2m}{r}}Hr^4]R_0^2+m^2 \mu \sech[\frac{2m^2 e^{-\frac{2m}{r}}}{r^4 R_0}]^2[e^{\frac{4m}{r}}r^8 R_0^2+32m^2(m-2r)^2\\
&[-2+3\sech[\frac{2m^2 e^{-\frac{2m}{r}}}{r^4 R_0}]^2+8e^{\frac{2m}{r}}r^4(3m^2-12mr+10r^2)R_0 \tanh[\frac{2m^2 e^{-\frac{2m}{r}}}{r^4 R_0}]]]
\end{split}
\end{align}
and
\begin{align}
\begin{split}
p_t={}&\frac{e^{-\frac{4m}{r}}}{r^8 R_0}[e^{\frac{2m}{r}}r^4[m^2+Hr^4 e^{\frac{2m}{r}}]R_0+m^2 \mu \sech[\frac{2m^2 e^{-\frac{2m}{r}}}{r^4 R_0}]^2[-e^{\frac{2m}{r}}r^4 R_0-\\
&8(m-2r)(m-r)\tanh[\frac{2m^2 e^{-\frac{2m}{r}}}{r^4 R_0}]]]
\end{split}
\end{align}
\par
\begin{figure}[H]
\begin{subfigure}[b]{0.4\textwidth}
\includegraphics[width=\textwidth]{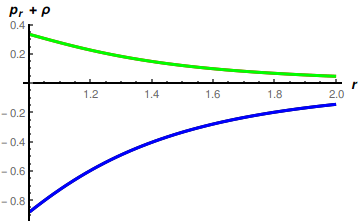}
\caption{\fcolorbox{green}{green}{\rule{0pt}{2.5pt}\rule{0pt}{2.5pt}} $\mu=+ve,R_0=+ve\;\;or\;\; \mu=+ve,R_0=-ve$, 
 \fcolorbox{blue}{blue}{\rule{0pt}{2.5pt}\rule{0pt}{2.5pt}} $\mu=-ve,R_0=+ve \;\;or\;\; \mu=-ve,R_0=-ve$}
\end{subfigure}
\hfill
\begin{subfigure}[b]{0.4\textwidth}
\includegraphics[width=\textwidth]{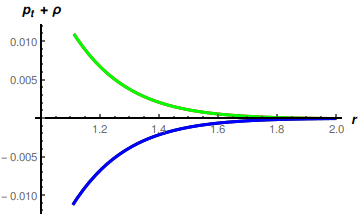} 
\caption{\fcolorbox{green}{green}{\rule{0pt}{2.5pt}\rule{0pt}{2.5pt}} $\mu=+ve,R_0=+ve \;\;or\;\; \mu=+ve,R_0=-ve$, 
 \fcolorbox{blue}{blue}{\rule{0pt}{2.5pt}\rule{0pt}{2.5pt}} $\mu=-ve,R_0=+ve \;\;or\;\; \mu=-ve,R_0=-ve$} 
\end{subfigure}
\hfill
\begin{subfigure}[b]{0.4\textwidth}
\includegraphics[width=\textwidth]{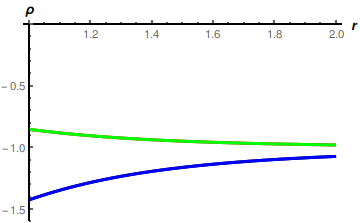} 
\caption{\fcolorbox{green}{green}{\rule{0pt}{2.5pt}\rule{0pt}{2.5pt}} $\mu=+ve,R_0=+ve \;\;or\;\; \mu=+ve,R_0=-ve$, 
 \fcolorbox{blue}{blue}{\rule{0pt}{2.5pt}\rule{0pt}{2.5pt}} $\mu=-ve,R_0=+ve \;\;or\;\; \mu=-ve,R_0=-ve$} 
\end{subfigure}
\hfill
\begin{subfigure}[b]{0.4\textwidth}
\includegraphics[width=\textwidth]{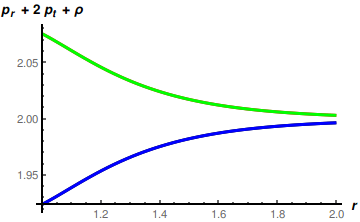} 
\caption{\fcolorbox{green}{green}{\rule{0pt}{2.5pt}\rule{0pt}{2.5pt}} $\mu=+ve,R_0=+ve \;\;or\;\; \mu=+ve,R_0=-ve$, 
 \fcolorbox{blue}{blue}{\rule{0pt}{2.5pt}\rule{0pt}{2.5pt}} $\mu=-ve,R_0=+ve \;\;or\;\; \mu=-ve,R_0=-ve$} 
\end{subfigure}
\hfill
\begin{subfigure}[b]{0.4\textwidth}
\includegraphics[width=\textwidth]{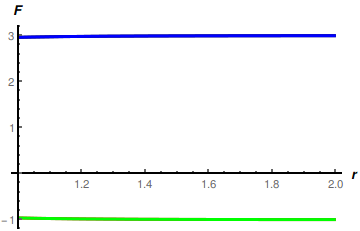} 
\caption{\fcolorbox{green}{green}{\rule{0pt}{2.5pt}\rule{0pt}{2.5pt}} $\mu=+ve,R_0=+ve \;\;or\;\; \mu=+ve,R_0=-ve$, 
 \fcolorbox{blue}{blue}{\rule{0pt}{2.5pt}\rule{0pt}{2.5pt}} $\mu=-ve,R_0=+ve \;\;or\;\; \mu=-ve,R_0=-ve$} 
\end{subfigure}
\caption{The plots depict $\rho+p_r$((a)), $\rho+p_t$((b)), $\rho$((c)), $\rho+p_r+2p_t$((d)) and $F=\frac{df}{dR}$((e)). All the energy conditions are partially violated for all combinations of $\mu$ and $R_0$}
\label{FIG.6}
\end{figure}

 Tsujikawa described that $\mu \in (0.905,1)$ \cite{tsujikawa2008observational} to sustain the viability of the model. Whereas for the violation of non-existence theorem of static spherically symmetric wormhole $F=\frac{df}{dR}>0$. We evaluated the geometric nature of wormhole structure through energy conditions for $\mu=1.0135$. From the following graphs FIG(\ref{FIG.6}) of $\rho$, $\rho+p_r$, $\rho+p_t$, $\rho+p_r+2p_t$ and $F=\frac{df}{dR}$, we get the analysis as,

\begin{itemize}
\item If $\mu=+ve$, $R_0=+ve$ or $\mu=+ve$, $R_0=-ve$, then $\rho+p_r \geq 0$.
\item If $\mu=+ve$, $R_0=+ve$ or $\mu=+ve$, $R_0=-ve$, then $\rho+p_t \geq 0$.
\item $\rho \leq$ for all combinations of $\mu$ and $R_0$.
\item $\rho+p_r+2p_t \geq 0$ for all combinations of $\mu$ and $R_0$.
\item $\frac{df}{dR}>0$ for $\mu=-ve, R_0=+ve$ or $\mu=-ve,R_0=-ve$ and $\frac{df}{dR}<0$ for $\mu=+ve, R_0=+ve$ or $\mu=+ve, R_0=-ve$.
\end{itemize}
We can conclude that, for $\mu=+ve$, $R_0=+ve$ or $\mu=+ve$, $R_0=-ve$ the NEC is respected by the exponential wormhole metric and $F=\frac{df}{dR}<0$, but at the same time WEC and SEC is partially violated as $\rho<0$. So for these particular combinations of $\mu$ and $R_0$, we get the wormhole solution which violates the non-existence theorem with the presence of minimal amount ofexotic matter.
\subsection{Gogoi-Goswami $f(R)$ gravity model}
\begin{figure}[H]
\begin{subfigure}[b]{0.4\textwidth}
\includegraphics[width=\textwidth]{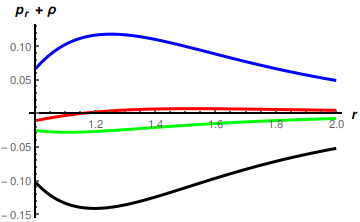}
\caption{\fcolorbox{red}{red}{\rule{0pt}{2.5pt}\rule{0pt}{2.5pt}} $a=+ve,R_0=+ve$, 
 \fcolorbox{black}{black}{\rule{0pt}{2.5pt}\rule{0pt}{2.5pt}} $a=-ve,R_0=+ve$, 
 \fcolorbox{green}{green}{\rule{0pt}{2.5pt}\rule{0pt}{2.5pt}} $a=+ve,R_0=-ve$, 
 \fcolorbox{blue}{blue}{\rule{0pt}{2.5pt}\rule{0pt}{2.5pt}} $a=-ve,R_0=-ve$}
\end{subfigure}
\hfill
\begin{subfigure}[b]{0.4\textwidth}
\includegraphics[width=\textwidth]{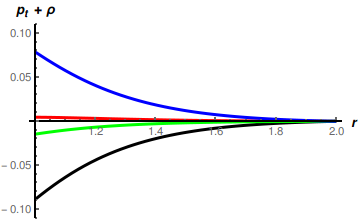} 
\caption{\fcolorbox{red}{red}{\rule{0pt}{2.5pt}\rule{0pt}{2.5pt}} $a=+ve,R_0=+ve$, 
 \fcolorbox{black}{black}{\rule{0pt}{2.5pt}\rule{0pt}{2.5pt}} $a=-ve,R_0=+ve$, 
 \fcolorbox{green}{green}{\rule{0pt}{2.5pt}\rule{0pt}{2.5pt}} $a=+ve,R_0=-ve$, 
 \fcolorbox{blue}{blue}{\rule{0pt}{2.5pt}\rule{0pt}{2.5pt}} $a=-ve,R_0=-ve$} 
\end{subfigure}
\hfill
\begin{subfigure}[b]{0.4\textwidth}
\includegraphics[width=\textwidth]{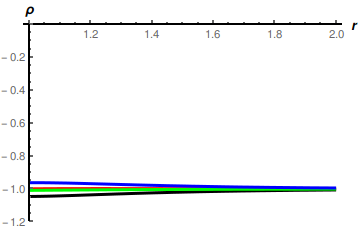} 
\caption{\fcolorbox{red}{red}{\rule{0pt}{2.5pt}\rule{0pt}{2.5pt}} $a=+ve,R_0=+ve$, 
 \fcolorbox{black}{black}{\rule{0pt}{2.5pt}\rule{0pt}{2.5pt}} $a=-ve,R_0=+ve$, 
 \fcolorbox{green}{green}{\rule{0pt}{2.5pt}\rule{0pt}{2.5pt}} $a=+ve,R_0=-ve$, 
 \fcolorbox{blue}{blue}{\rule{0pt}{2.5pt}\rule{0pt}{2.5pt}} $a=-ve,R_0=-ve$} 
\end{subfigure}
\hfill
\begin{subfigure}[b]{0.4\textwidth}
\includegraphics[width=\textwidth]{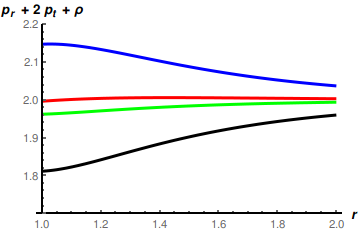} 
\caption{\fcolorbox{red}{red}{\rule{0pt}{2.5pt}\rule{0pt}{2.5pt}} $a=+ve,R_0=+ve$, 
 \fcolorbox{black}{black}{\rule{0pt}{2.5pt}\rule{0pt}{2.5pt}} $a=-ve,R_0=+ve$, 
 \fcolorbox{green}{green}{\rule{0pt}{2.5pt}\rule{0pt}{2.5pt}} $a=+ve,R_0=-ve$, 
 \fcolorbox{blue}{blue}{\rule{0pt}{2.5pt}\rule{0pt}{2.5pt}} $a=-ve,R_0=-ve$} 
\end{subfigure}
\hfill
\begin{subfigure}[b]{0.4\textwidth}
\includegraphics[width=\textwidth]{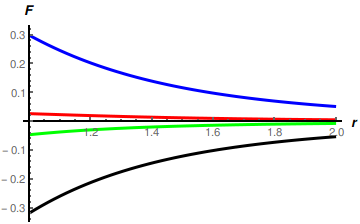} 
\caption{\fcolorbox{red}{red}{\rule{0pt}{2.5pt}\rule{0pt}{2.5pt}} $a=+ve,R_0=+ve$, 
 \fcolorbox{black}{black}{\rule{0pt}{2.5pt}\rule{0pt}{2.5pt}} $a=-ve,R_0=+ve$, 
 \fcolorbox{green}{green}{\rule{0pt}{2.5pt}\rule{0pt}{2.5pt}} $a=+ve,R_0=-ve$, 
 \fcolorbox{blue}{blue}{\rule{0pt}{2.5pt}\rule{0pt}{2.5pt}} $a=-ve,R_0=-ve$} 
\end{subfigure}
\caption{The plots depict $\rho+p_r$((a)), $\rho+p_t$((b)), $\rho$((c)), $\rho+p_r+2p_t$((d)) and $F=\frac{df}{dR}$((e)) for $\mu=+ve$. It shows that NEC is respected throughout the space-time for some specific combination of $a$ and $R_0$ while there is a partial violation of WEC and SEC.}
\label{FIG.7}
\end{figure}

\begin{figure}[H]
\begin{subfigure}[b]{0.4\textwidth}
\includegraphics[width=\textwidth]{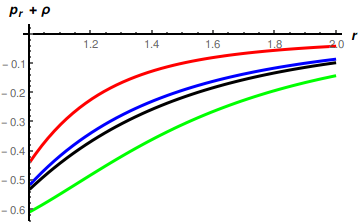}
\caption{\fcolorbox{red}{red}{\rule{0pt}{2.5pt}\rule{0pt}{2.5pt}} $a=+ve,R_0=+ve$, 
 \fcolorbox{black}{black}{\rule{0pt}{2.5pt}\rule{0pt}{2.5pt}} $a=-ve,R_0=+ve$, 
 \fcolorbox{green}{green}{\rule{0pt}{2.5pt}\rule{0pt}{2.5pt}} $a=+ve,R_0=-ve$, 
 \fcolorbox{blue}{blue}{\rule{0pt}{2.5pt}\rule{0pt}{2.5pt}} $a=-ve,R_0=-ve$}
\end{subfigure}
\hfill
\begin{subfigure}[b]{0.4\textwidth}
\includegraphics[width=\textwidth]{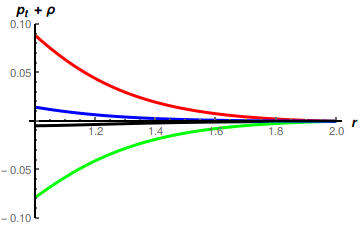} 
\caption{\fcolorbox{red}{red}{\rule{0pt}{2.5pt}\rule{0pt}{2.5pt}} $a=+ve,R_0=+ve$, 
 \fcolorbox{black}{black}{\rule{0pt}{2.5pt}\rule{0pt}{2.5pt}} $a=-ve,R_0=+ve$, 
 \fcolorbox{green}{green}{\rule{0pt}{2.5pt}\rule{0pt}{2.5pt}} $a=+ve,R_0=-ve$, 
 \fcolorbox{blue}{blue}{\rule{0pt}{2.5pt}\rule{0pt}{2.5pt}} $a=-ve,R_0=-ve$} 
\end{subfigure}
\hfill
\begin{subfigure}[b]{0.4\textwidth}
\includegraphics[width=\textwidth]{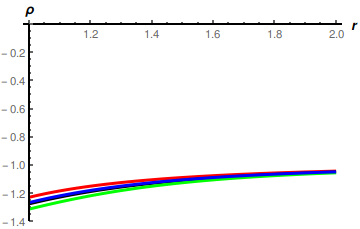} 
\caption{\fcolorbox{red}{red}{\rule{0pt}{2.5pt}\rule{0pt}{2.5pt}} $a=+ve,R_0=+ve$, 
 \fcolorbox{black}{black}{\rule{0pt}{2.5pt}\rule{0pt}{2.5pt}} $a=-ve,R_0=+ve$, 
 \fcolorbox{green}{green}{\rule{0pt}{2.5pt}\rule{0pt}{2.5pt}} $a=+ve,R_0=-ve$, 
 \fcolorbox{blue}{blue}{\rule{0pt}{2.5pt}\rule{0pt}{2.5pt}} $a=-ve,R_0=-ve$} 
\end{subfigure}
\hfill
\begin{subfigure}[b]{0.4\textwidth}
\includegraphics[width=\textwidth]{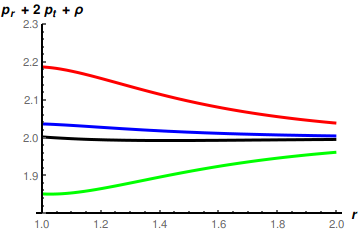} 
\caption{\fcolorbox{red}{red}{\rule{0pt}{2.5pt}\rule{0pt}{2.5pt}} $a=+ve,R_0=+ve$, 
 \fcolorbox{black}{black}{\rule{0pt}{2.5pt}\rule{0pt}{2.5pt}} $a=-ve,R_0=+ve$, 
 \fcolorbox{green}{green}{\rule{0pt}{2.5pt}\rule{0pt}{2.5pt}} $a=+ve,R_0=-ve$, 
 \fcolorbox{blue}{blue}{\rule{0pt}{2.5pt}\rule{0pt}{2.5pt}} $a=-ve,R_0=-ve$} 
\end{subfigure}
\hfill
\begin{subfigure}[b]{0.4\textwidth}
\includegraphics[width=\textwidth]{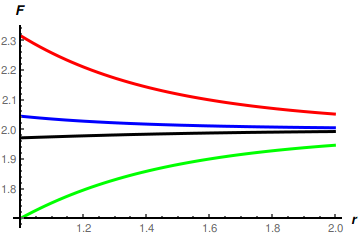} 
\caption{\fcolorbox{red}{red}{\rule{0pt}{2.5pt}\rule{0pt}{2.5pt}} $a=+ve,R_0=+ve$, 
 \fcolorbox{black}{black}{\rule{0pt}{2.5pt}\rule{0pt}{2.5pt}} $a=-ve,R_0=+ve$, 
 \fcolorbox{green}{green}{\rule{0pt}{2.5pt}\rule{0pt}{2.5pt}} $a=+ve,R_0=-ve$, 
 \fcolorbox{blue}{blue}{\rule{0pt}{2.5pt}\rule{0pt}{2.5pt}} $a=-ve,R_0=-ve$} 
\end{subfigure}
\caption{The plots depict $\rho+p_r$((a)), $\rho+p_t$((b)), $\rho$((c)), $\rho+p_r+2p_t$((d)) and $F=\frac{df}{dR}$((e)) for $\mu=-ve$. All the energy conditions are partially violated for all possible combinations of $a$ and $R_0$.}
\label{FIG.8}
\end{figure}

It is a new viable $f(R)$ gravity model, constructed by Gogoi and Goswami \cite{gogoi2020new}. This model is defined as,
\begin{equation}
f(R)=R-\frac{a}{\pi}R_0 \cot^{-1}(\frac{R_0^2}{R^2})-\mu R_0[1-e^{-\frac{R}{R_0}}],
\end{equation}
where $a$ and $\mu$ are two dimensionless constants and $R_0$ is a characteristic curvature constant having dimensions same as curvature scalar $R$. The allowed range for $a$ is $-1.68381<a<0.367545$. Now from the plots FIG(\ref{FIG.7}) and FIG(\ref{FIG.8}) of $\rho$, $\rho+p_r$, $\rho+p_t$, $\rho+p_r+2p_t$ and $F=\frac{df}{dR}$, we get the following analysis,

\par If $\mu =+ve$
\begin{itemize}
\item If $a=-ve$, $R_0=-ve$, then $\rho+p_r \geq 0$ (also for $a=+ve$, $R_0=+ve$ if $r>1.55$).
\item If $a=+ve$, $R_0=+ve$ or $a=-ve$, $R_0=-ve$, then $\rho+p_t \geq 0$.
\item $\rho \leq 0$, for all combinations of $a$ and $R_0$.
\item $\rho+ p_r+2p_t \geq 0$, for all combinations of $a$ and $R_0$. 
\item $\frac{df}{dR}>0$ for $a=+ve, R_0=+ve$ or $a=-ve, R_0=-ve$ and $\frac{df}{dR}<0$ for $a=+ve, R_0=-ve$ or $a=-ve, R_0=+ve$.
\end{itemize}
\par If $\mu =-ve$
\begin{itemize}
\item $\rho+p_r \leq 0$, for all combinations of $a$ and $R_0$.
\item If $a=+ve$, $R_0=+ve$ or $a=-ve$, $R_0=-ve$, then $\rho+p_t \geq 0$.
\item $\rho \leq 0$, for all combinations of $a$ and $R_0$.
\item $\rho+ p_r+2p_t \geq 0$, for all combinations of $a$ and $R_0$.
\item $\frac{df}{dR}>0$ for all combinations of $a$ and $R_0$.
\end{itemize}
From these we can conclude that, NEC is respected in the exponential wormhole geometry only if $\mu=+ve$,  $a=-ve$, $R_0=-ve$. But the WEC and the SEC are partially violated throughout the space-time, which represents that the wormhole structure has a normal matter at the throat. Again $F=\frac{df}{dR}>0$, indicates the non-spherical symmetry of the wormhole solution. Similar results can be achieved for $\mu=+ve$, $a=+ve$, $R_0=+ve$ but only if $r>1.55$. That is for this specific combination, NEC is violated at the throat which shows that the wormhole contains a small amount of exotic matter near the throat.
\section{Results and Discussion}
\par In this paper, we have carried out a comparative study of the so called "exponential" wormhole metric in General Relativity and modified $f(R)$ theory of gravity. We have constructed the field equations for this exponential metric for both the cases. In recent years, many researchers studied the Morris-Thorne wormhole with different redshift and shape function in various viable modified theory of gravity, but no one has ever studied the exponential wormhole metric in $f(R)$ gravity.
\par The radius of the throat comes out as $r=m$, all the metric components are finite and the diagonal components are non-zero at $r=m$. In order to be class one (according to Karamarkar condition), $m$ can take values $m=0$ or $m=r$ or $m=\frac{3}{2}r$. Out of which $m=0$ is forbidden due to the flare-out condition. Again $m=r$ represents the throat radius, so the exponential metric acts as a class one static and spherically symmetric line element at the throat. We have also studied this exponential wormhole metric in four viable $f(R)$ gravity model, namely exponential model, Starobinsky model, Tsujikawa model and Gogoi-Goswami $f(R)$ gravity model. The results obtained from the study in General Relativity and those viable $f(R)$ gravity models are as follows,
\par in General Relativity,
\begin{itemize}
\item From the field equations in General Relativity, we have examined the energy conditions and it is found that $\rho+p_r<0$, $\rho+p_t=0$, $\rho<0$ and $\rho+p_r+2p_t=0$, that is the NEC, WEC and SEC are violated throughout the space-time indicating the presence of exotic matter.
\item The exponential wormhole metric obeys the flare-out condition everywhere and it doesnot possess any kind of singularity.
\item All the curvature components and the scalar invariants are finite everywhere, they are finite at the throat and decay to zero as $r\to \infty$ and as $r\to 0$.
\item The exponential wormhole metric violates the null Ricci Convergence condition which is important for the better understanding of the flare-out conditions.
\item The exponetial wormhole metric gives the scalar field equation of motion $(g^{ab}\nabla_a \nabla_b)\Phi=0$ and we can derive Einstein equation for a negative kinetic energy massless scalar field or phantom field, which is a evidence that the exponential wormhole metric represents a traversable wormhole
\end{itemize}
 In modified $f(R)$ theory of gravity, we have obtained the field equations and also studied the energy conditions in four viable $f(R)$ gravity model. The function $f(R)$ has some free parameters or constants in these viable models. Some specific combinations of these parameters/constants show some interesting results in energy conditions which are quite different from those in General Relativity.
\begin{itemize}
\item In case of Exponential $f(R)$ gravity model, if we consider $\mu=+ve$, $R_0=-ve$, then $\rho+p_r \geq 0$, $\rho+p_t \geq 0$ and $\rho+p_r+2p_t \geq 0$ (all possible combinations) but $\rho \leq 0$ (for all possible combinations). We can conclude that in case of exponential $f(R)$ gravity model, the exponential wormhole metric obeys the necessary NEC and $F=\frac{df}{dR}<0$ but partially violates the WEC and SEC. So $\mu=+ve, R_0=+ve$ is the perfect combination in order to get wormhole solution which violates the non-existence theorem with the presence of insignificant amount of exotic matter.
\item In case of Starobinsky $f(R)$ gravity model, if $l=+ve$, $a=+ve$, $R_0=+ve$ or $l=+ve$, $a=-ve$, $R_0=-ve$ (for $r>1.2$) or $l=-ve$, $a=-ve$, $R_0=+ve$ or $l=-ve$, $a=+ve$, $R_0=-ve$ (for $r>1.33$), then $\rho+p_r \geq 0$, $\rho+p_t \geq 0$, $\rho+p_r+2 p_t \geq 0$ (for all possible combinations) but $\rho \leq 0$ (for all possible combinations). So for these combinations of $l$, $a$ and $R_0$, NEC is respected outside the throat and $F=\frac{df}{dR}>0$, which signifies that the wormhole has a non-spherical symmetry and the throat is filled with a small amount of exotic matter.
\item In case of Tsujikawa $f(R)$ gravity model, if $\mu=+ve$, $R_0=+ve$ or $\mu=+ve$, $R_0=-ve$, then $\rho+p_r \geq 0$, $F=\frac{df}{dR}<0$ and $\rho+p_t\geq 0$, $\rho+p_r+2p_t \geq 0$ (for all possible combinations) but for all possible combinations of $\mu$ and $R_0$, $\rho \leq 0$. While WEC and SEC is violated throughout the space-time. So $\mu=+ve$, $R_0=+ve$ or $\mu=+ve$, $R_0=-ve$ are the perfect combinations to get traversable wormhole in Tsujikawa $f(R)$ gravity model with negligible amount of exotic matter.
\item In case of Gogoi-Goswami $f(R)$ gravity model, $\rho+p_r$ and $\rho+p_t\geq0$ for $\mu=+ve$, $a=-ve$, $R_0=-ve$. But $\rho+p_r+2p_t \geq 0$ and $\rho \leq 0$ for all possible combinations of $\mu$, $a$ and $R_0$. So NEC is respected for $\mu=+ve, a=-ve, R_0=-ve$, while WEC and SEC are violated for all possible combinations. Again for these specific combinations of $\mu$, $a$ and $R_0$, $F=\frac{df}{dR}>0$, which implies the non-spherical symmetry of the wormhole with the normal matter present at the throat. Again if $\mu=+ve$, $a=+ve$, $R_0=+ve$, the wormhole again has non-spherical symmetry but this time the throat contains the exotic matter. While the NEC is respected just outside the wormhole throat. 
\end{itemize}
Thus in comparison with General Relativity, the exponential wormhole metric obeys the necessary NEC at the throat in modified $f(R)$ gravity model for some particular combinations of the free parameters$/$constants. So the exponential wormhole metric could form a traversable wormhole geometry with negligible amount of exotic matter.
\section*{Acknowledgement}
This work is supported by University Grants Commission, Ministry of Education, Govt. of India(NFOBC No.F. 82-44/2020(SA-III)) under the scheme NFOBC programme.
\\
\bibliographystyle{ieeetr}
\bibliography{myref}

\end{document}